\documentclass{PoS}
\newcommand{\nc}{\newcommand}
\nc{\beq}{\begin{equation}}
\nc{\eeq}{\end{equation}}
\nc{\beqa}{\begin{eqnarray}}
\nc{\eeqa}{\end{eqnarray}}
\def\be{\begin{equation}}
\def\ee{\end{equation}}    
\newcommand{\da}{\dot{a}}
\newcommand{\db}{\dot{b}}

\newcommand{\dda}{\ddot{a}}

\newcommand{\pa}{a^{\prime}}

\newcommand{\pn}{n^{\prime}}
\newcommand{\ppa}{a^{\prime \prime}}

\newcommand{\ppn}{n^{\prime \prime}}
\newcommand{\fda}{\frac{\da}{a}}

\newcommand{\fdb}{\frac{\db}{b}}

\newcommand{\fdda}{\frac{\dda}{a}}

\newcommand{\fpa}{\frac{\pa}{a}}

\newcommand{\fpn}{\frac{\pn}{n}}
\newcommand{\fppa}{\frac{\ppa}{a}}

\newcommand{\fppn}{\frac{\ppn}{n}}

\title{Braneworld Cosmology}

\ShortTitle{Braneworld Cosmology}

\author{\speaker{James M.\ Cline}\\
        McGill Univeristy, Montr\'eal, Qc H3A2T8, Canada\\
        E-mail: \email{jcline@physics.mcgill.ca}}

\abstract{A brief review of the field of braneworld cosmology, 
from its inception with the large extra dimension scenario,
to aspects of cosmology in warped extra dimensions,
including the RS-I and RS-II models, braneworld inflation, 
the Goldberger-Wise mechanism, mirage cosmology, 
the radion-induced phase transition in RS-I, possible gravity 
wave signals,  and the DGP model.}

\FullConference{From Strings to LHC\\
		January 2-10 2007\\
		International Centre Dona Paula, Goa India}

\begin{document}

I was asked by the organizers of ``Strings to LHC'' to provide a
brief overview of brane cosmology.  This is a large subject, and my
scope will be necessarily limited to some of the key ideas.  In
particular, I will not discuss the many interesting developments with
higher-codimension branes, rather focusing on the simplest case of
codimension-one.   Furthermore, this is not a review of string
cosmology, nor of brane-antibrane inflation.   Braneworld cosmology
is a string-inspired framework, where the ``branes'' are merely
delta-function sources in the stress-energy tensor, which generally
do not have the other properties of D-branes, such as charge and
internal dynamics.    I will give a historically ordered account
starting with large extra dimensions, then turning to warped
geometries and the DGP model.  

\section{Flat extra dimensions}

The modern era of brane cosmology began with ref.\ \cite{ADD} in the
context of large extra dimensions, made possible by the hypothesis
that the standard model of particle physics is localized on a
D-brane.  Since only gravity was presumed to propagate in the  extra
dimensions, only short-distance (sub-millimeter) probes of the
gravitational force could constrain the size of the extra dimensions.
Experimentally it was known that gravity did not deviate from
its known four-dimensional form down to distances just below a
millimeter, hence allowing for extra dimensions of that magnitude.
Additionally, with two extra dimensions, the fundamental gravity scale
$M_d$ would be lowered to a value consistent with the TeV scale, using the
relation  
\be
	M_p^2 \sim M_d^{d-2} R^{d-4}
\ee
in $d$ spacetime dimensions, with $d-4$ having size $R$.  Cosmology
becomes highly constrained in such a scenario because the phase space
for Kaluza-Klein gravitons $\tilde g$ is so large that they would be
overproduced if the temperature of the universe was ever greater than
some maximum value $T_* \sim 1$ MeV.  The gravitons are populated by
interactions like $\gamma\gamma\to \tilde g \tilde g$.  Even though
each interaction vertex is suppressed by $1/M_p$ in the 4D effective
theory, the phase space is so dense that the total cross section is
similar to having couplings that are only suppressed by $1/M_6$, where
$M_6$ at the $1-10$ TeV scale.  The KK gravitons are constrained because
they can overclose the universe, or distort the diffuse $\gamma$ ray 
spectrum by their decays at late times.

One of the key papers motivating interest in brane cosmology was
ref.\ \cite{BDL} (BDL).  Their work was inspired by string theory,
the Ho\v rava-Witten model \cite{HW}, in which the 11th dimension of
heterotic M-theory is compactified on an interval $y\in[0,1]$, which
can be considered as a circle parametrized by $y\in [-1,1]$ with
points identified under a $Z_2$ orbifold symmetry $y\to -y$. The ends
of the interval are the orbifold fixed points.   BDL considered a 5D
line element of the form 
\be 
	ds^2 = -n^2(y) dt^2 + a^2(t,y) d\vec x^2 + b_0^2 dy^2
\ee
with branes at $y=0$ and $y=1$, and $b_0$ assumed to be constant.

The branes provide delta-function sources in the Einstein equations,
which take the simple form
\beqa
&& H^2 = \left( \fda\right)^2 = \frac{n^2}{b_0^2}
\left(\fppa + \left( \fpa \right)^2 \right) + {n^2\over 3 M_5^3 b_0} 
\sum_i \rho_i \delta(y-y_i), \label{ein00}\\
&&\left(\fda\right)^2 +2\fdda  =
\frac{n^2}{b_0^2} \left\{ 2\fppa 
 +\fppn +
\fpa\left(\fpa+2\fpn\right) 
\right\}  - {n^2\over 
M_5^3 b_0}\sum_i p_i \delta(y-y_i)
\label{einii} \\
&&\left(\fda\right)^2 + \fdda =  \frac{n^2}{b_0^2}
\fpa \left(\fpa+\fpn \right) 
\label{ein55}\\
&&\fpn \fda + \fpa \fdb - \frac{\dot{a}^{\prime}}{a}
= 0 
\label{ein05}
\end{eqnarray}
For constant equation of state $p/\rho = w$ on both branes, the
solution is also simple,
\beqa
\label{soln}
	a &=& t^q\left(1 - {b_0\rho_0\over 6 M_5^3}|y|\right)\\
	n &=& 1 +  {b_0\rho_0\over 2 M_5^3}\left(w + \frac23 \right)|y|
\eeqa
where $q^{-1} = 3(1+w)$.  The time-dependence of the scale factor
differs from that in standard cosmology, where $q^{-1} = 3(1+w)/2$.
A simple way of seeing why this happens is to consider the Einstein
equation (\ref{ein00}).  By integrating it over the region
$y\in[-\epsilon,\epsilon]$, we find that
\beq
\label{j1}
	 \int_{-\epsilon}^\epsilon{a''\over a} dy
 = {\Delta a'\over a} = 2{a'\over a} = - {b_0\rho \over 3 M_5^2}
\eeq
where we used $a'(-\epsilon) = -a'(\epsilon)$ due to 
the $Z_2$ orbifold symmetry.  Similarly, integrating (\ref{einii})
in this region gives
\beq
\label{j2}
	 {n'\over n} = {b_0\over 3 M_5^2}(3 p - \rho)
\eeq 
Using these values in (\ref{ein55}) and choosing $n=1$ at $y=0$
gives
\beq
\label{feq2}
	H^2 + \fdda = {1\over 36 M_5^6}\rho(\rho- 3p)
\eeq
Because of energy conservation, $\dot\rho = 3H(\rho+p)$, which follows
from (\ref{ein05}), eq.\ (\ref{feq2}) is consistent with the first-order equation
\beq
\label{sfe}
	H^2 = \left({\rho\over 6 M_5^3}\right)^2
\eeq
This modified form for the Friedmann equation can be solved for $a(t)$
given the usual scaling $\rho \sim a^{-3(1+w)}$, to show that
$a(t)\sim t^q$ as in (\ref{soln}).  

Needless to say, the dependence $H\sim \rho$ instead of the
standard relation $H = \sqrt{8\pi G\rho/3}$ is ruled out by the
successful prediction of big bang nucleosynthesis (BBN) for the light
element abundances.  BBN is quite sensitive to changes in the
expansion rate of the universe.  

\section{Warped extra dimensions}

To overcome the problem with BDL's Friedmann equation, a simple
modification was proposed in \cite{CGKT} and \cite{CGS}: add a
negative bulk cosmological constant, $\Lambda_b$, which further
modifies the Friedmann equation to the form
\beq
	H^2 = \left({\rho\over 6 M_5^3}\right)^2 + {\Lambda_b\over
	6 M_5^3}
\eeq
This greatly improves the situation since we can let
\beq
	\rho = \tau + \rho_m
\eeq
where $\tau$ represents the tension of the brane at $y=0$, while $\rho_m$ stands for
ordinary matter or radiation on the brane.  By tuning $\tau\to\tau_0$
in such a way  that the
constant contribution to $H$ vanishes,
\beq
	\left({\tau_0\over 6 M_5^3}\right)^2 + {\Lambda_b\over
	6 M_5^3} = 0
\eeq
which is the usual tuning of the cosmological constant to zero,
we are left with the modified Friedmann equation
\beq
\label{mfe}
	H^2 = {\tau\over 18 M_5^3}\rho_m \left(1 + {\rho_m\over
	2\tau_0}\right)
\eeq
where we should identify the coefficient as
\beq
	{\tau_0\over 18 M_5^3} = {8\pi\over 3} G
\eeq
This is interesting because it reproduces the standard 4D result at
low energies $\rho_m\ll \tau_0$, but predicts the BDL-like modification
$H\sim \rho$ in the high-energy regime, $\rho_m\gg\tau_0$.

Moreover, the background solution in the presence of $\Lambda_b<0$
becomes the warped solution of Randall and Sundrum (RS) \cite{RSI}
in the limit $\rho_m\to 0$,
\beq
	a(y) = n(y) = e^{-bk|y|}  
\eeq
thus providing a link between brane cosmology and the warped
compactification scenario which was proposed for completely
noncosmological reasons---namely to solve the hierarchy problem.
(If $\tau$ is not tuned to the static value $\tau_0$, one finds
warped de Sitter solutions $a(t,y) = e^{Ht}(\cosh bk|y| - 
(\tau/\tau_0)\sinh bk|y|)$ \cite{Kaloper}.) 

However this was not a complete solution to the problems which plagued
the BDL model.  In both BDL and in RS it was necessary to have a
negative tension brane at $y=1$, with equal and opposite tension to
the $y=0$ brane.  This is not fatal since within string theory there
exist physically consistent negative-tension objects, orientifold
planes.  The real problem was that the matter densities on the two
branes had to also be tuned, such that
\beq
\label{tuning}
	\rho_1 = - e^{2kb_0} \rho_0
\eeq
where the subscript indicates the position of the brane.  String
theory does not provide any sources of {\it matter} with negative energy
density.  Furthermore, it was not clear why it should be necessary to
tune the matter densities on the two branes.  

In retrospect, it should have been obvious that something else was
seriously lacking in the model of \cite{BDL} since  it failed to
reproduce the predictions of general relativity (GR) even if the
extra dimension became arbitrarily small---notice that the strange
Friedmann equation (\ref{sfe}) is completely insensitive to the value
of $b_0$, which is the size of the extra dimension. The problem was
shown to be due to artificially assuming that $b$ was static, even
though there was no mechanism in the theory to stabilize it at any
particular value \cite{CGRT}.  This criticism applied equally to the
warped models of \cite{CGKT, CGS}.

In fact, $b_0$ should be treated as a massless scalar field
$b(x^\mu)$, the {\it radion}, which leads to a scalar-tensor theory of
gravity rather than GR.  Although it is possible to find solutions
where this massless scalar is constant by tuning the brane
matter densities as in (\ref{tuning}), fifth force constraints oblige
us to give the radion a mass, and stabilize it at some particular
value.  Ref.\  \cite{CGRT} showed that once this is done, the standard
Friedmann equation is recovered in either the warped or unwarped
models, up to corrections which are suppressed by the radion mass,
rather than the brane tension.  

It is interesting to see how the argument which led to eq.\
(\ref{sfe}) is changed once we add a potential energy $V(b)$
for the radion (and possibly a bulk cosmological constant 
$\Lambda_b$).
The 5D Einstein equations become
\beqa
 3H^2 - 3\frac{n^2}{b_0^2}
\left(\fppa + \left( \fpa \right)^2 \right) &=& {n^2\over M_5^3} 
\left( \sum_i \rho_i \delta(b_0(y-y_i))
+ \Lambda_b + V(b_0)\right)
 \label{ein00a}\\
\!\!\!\!\!\!\!\!\!\!\!\!\!\!H^2 +2\fdda  
-\frac{n^2}{b_0^2} \left\{ 2\fppa  +\fppn +
\fpa\left(\fpa+2\fpn\right) \right\}  
&=& - {n^2\over 
M_5^3}\left(\sum_i p_i \delta(b_0(y-y_i)) +  \Lambda_b + V(b_0)\right)
\label{einiia} \\
-3\left(H^2 + \fdda\right) + 3\frac{n^2}{b_0^2}
\fpa \left(\fpa+\fpn \right) &=& - {n^2\over M_5^3}
\Bigg(\Lambda_b + V(b_0) + b_0 V'(b_0)\Bigg)
\label{ein55a}
\end{eqnarray}
where $V(b_0)$ is the radion potential, and we are still assuming that
$b_0$ is constant in space and time for simplicity.
The junction conditions (\ref{j1}-\ref{j2}) are unchanged by the
radion potential, but the 5-5 Einstein equation (\ref{ein55a}) no
longer determines $H^2 + \fdda$ in terms of brane sources because
of the extra $V(b_0)$ contribution.  Instead this equation determines
the shift in the radion due to the expansion of the universe,
$b\to b_0 + \delta b$.
For the Randall-Sundrum model, ref.\ \cite{CGRT} finds
\beq
\label{db}
	{\delta b\over b} \sim {(\rho-3p)\over m_b^2 a^2 M_p^2}
\eeq
where $m_b$ is the radion mass, and $a$ is the warp factor; therefore
$a M_p\equiv \Lambda$ is the TeV (or infrared) scale. 

It is no longer simple to compute the modifications to the Friedmann
equation once the radion is stabilized.  Moreover the result depends
upon whether the expansion of the universe is measured at the Planck
brane ($y=0$) or the TeV brane $(y=1)$.  At the TeV brane, the leading correction was 
found to be \cite{CV}
\beq
	\left.H^2\right|_{y=1} = {8\pi G\over 3}
	\left(\rho + {c\rho^2\over m_b^2\Lambda^2}(1-3w)(1+w)
	+O(\rho^2)\right)
\eeq
where $c = O(1)$, $\rho$ is the warped energy density on the TeV brane, and 
$w=p/\rho$.  Curiously the $\rho^2$ correction vanishes when $w=1/3$
(radiation) or $w=-1$ (inflation).  

\section{Randall-Sundrum II cosmology}

Despite its shortcomings (because it ignored the need for radion
stabilization), the modified Friedmann equation (\ref{mfe}) turned out
to have a deeper significance.  This is in connection with the 
Randall-Sundrum II model \cite{RSII}, in which the TeV brane is
displaced to $y\to\infty$ (equivalently $b_0\to\infty$) and thereby removed from the
theory.  There is no physical radion in this theory (it can be gauged
away by a coordinate transformation), hence no need for
stabilization.  Eq.\ (\ref{mfe}) is still the correct Friedmann
equation, but now it is fully justified, unlike in the RSI model.
Big bang nucleosynthesis puts a weak constraint on the brane tension;
$\tau > $ (MeV)$^4$ to insure that $H$ has the standard form during
BBN.  Terrestial measurements of the gravitational force give a much
stronger constraint since the exchange of KK gravitons 
modifies the Newtonian gravitational potential to the form
\beq
	V = -G{m_1 m_2\over r}\left(1 + {1\over r^2 k^2}\right)
\eeq
with $k = M_5^3/M_p^2$.  Interpreting the current limit from the
E\"ot-Wash experiment \cite{eotwash} to imply that $kr > 1$ at 
the scale 0.04 mm, we find the bound
\beq
	k > 5 {\rm\ meV}
\eeq
hence
\beq
	M_5 = k^{1/3} M_p^{2/3} > 3\times 10^5 {\rm \ TeV}
\eeq
and the brane tension is
\beq
	\tau = 24 k M_5^3 > ( 7 {\rm\ TeV} )^4,
\eeq
far above the energy scale of BBN.

\subsection{Inflation with nonstandard Friedmann equation}

An obvious application of the modified Friedmann equation is to
inflation at scales greater than $\tau^{1/4}$, where $H\sim\rho$.
Ref.\ \cite{MWBH} examined chaotic inflation in this context, noting
that one can get inflation to work with steeper potentials than 
usual, since $H$ is larger than its conventional value and Hubble
damping is more effective.  It was shown that the slow-roll parameters
get modified relative to their conventional values (denoted by
$\epsilon_0$ and $\eta_0$),
\beqa
	\epsilon &=& \epsilon_0\left({1 + V/\tau\over (1+V/2\tau)^2}
	\right) \sim {4\tau\over V}\epsilon_0\\
	\eta &=& \eta{1\over 1+V/2\tau} \sim {2 \tau\over V}\eta_0
\eeqa
where $V$ is the potential and the final estimates are in the limit
$V\gg\tau$.  One can show that inflation occurs when $\dot\phi^2
< \frac25 V$ ($p < -\frac23\rho$)
intead of the usual condition $\dot\phi^2 < V$ ($p< -\frac13\rho$) 
in this
regime.  Furthermore the number of e-foldings as a function of the
inflaton field is given by
\beq
	N \cong -{1\over M_p^2}\int d\phi {V\over V'}
	\left(1 + {V\over 2\tau}\right)
\eeq
and the scalar power amplitude is enhanced relative to its usual
value,
\beq
	A_s^2 = A_0^2 \left(1 + {V\over 2\tau}\right)^2
\eeq
The spectral index formula keeps its usual form in the new slow-roll
parameters, but their relation to the number of e-foldings 
in general differs
from the standard result (shown below in parentheses)
 in the high-energy regime \cite{LS}:
\beq
	n_s -1 \cong -6\epsilon+2\eta \cong\left\{
	\begin{array}{ll} -\frac{5}{2N}\hbox{\ (c.f.\ }-\frac{2}{N}),
	& V = m^2\phi^2\\
	-\frac{3}{N} \hbox{\ \ \,(c.f.\ }-\frac{3}{N}),&
	V = \lambda\phi^4\end{array}\right.
\eeq
Coincidentally, the prediction for $\phi^4$ chaotic inflation does not
change, although \cite{LS} shows that in the intermediate region where
$V\sim\tau$ this is no longer the case, and the tension between
$\lambda\phi^4$
theory and the CMB data is made somewhat worse.  For the $\phi^2$
model, the increase in $|n_s-1|$ in the high $V$ regime degrades the
agreement with the data (viewed as likelihood contours in the plane of
the tensor-to-scalar ratio $r$ versus $n_s-1$) relative to the standard
$V\ll\tau$ regime: it pushes it outside the 1-$\sigma$ contour.

A novel feature is that inflation can occur with subPlanckian field
values, $\phi < M_p$, which is not the case for conventional chaotic
inflation, and is often regarded as a drawback.  However, one still
needs $\phi > M_5$; for $V=m^2\phi^2$, inflation occurs for $\phi\sim
300 M_5$, and the COBE normalization requires $m = 5\times 10^{-5}
M_5$.  

Braneworld inflation was also considered using the potential
\beq
\label{vexp}
	V = V_0 e^{-\alpha\phi/M_p}
\eeq
in \cite{CLL}, which relies upon gravitational particle production to 
get reheating since there is no minimum around which $\phi$ can
oscillate to give ordinary reheating.  In standard cosmology, one
needs to have $\alpha^2 < 2$ in order to get inflation from
(\ref{vexp}), but this is no longer true in the high-energy regime.
Ref.\ \cite{CLL} finds that with the COBE normalization the
inflationary scale, brane tension and 5D Planck mass are fixed in
terms of $\alpha$ as
\beq
	V_{\rm inf} \sim \left({10^{15}{\rm\ GeV}\over
\alpha}\right)^4,\qquad \tau \sim \left({10^{15}{\rm\ GeV}\over
\alpha^{3/2}}\right)^4,\qquad M_5\sim 
{3\times 10^{16}{\rm\ GeV}\over \alpha}
\eeq
The number of e-foldings as a function of field value is given by
\beq
	N = {1\over 2\tau\alpha^2}\left(
	V(\phi) - V_{\rm end}\right)
\eeq
in the high-energy phase.  This shows that $\alpha$ can be taken much
larger than the standard range if $\tau$ is appropriately adjusted 
downward, and so the potential can be much steeper than usual while
still supporting slow-roll inflation.  The eventual reheating temperature is found to be
$T \sim 10^3/\alpha$ GeV.  Independently of $\alpha$, the predicted
relation between spectral index and number of e-foldings is 
\beq
	n_s -1 = -{4\over N+1}
\eeq
leading to $n_s \cong 0.92$ as well as the
tensor-to-scalar ratio $r=0.03$.  This value of $n_s$ is somewhat too
low relative to the currently favored WMAP3 value $n_s=0.95$.

\section{Randall-Sundrum I cosmology}
We now return to the more rich and complicated case of two
branes in a compact warped geometry.  Notice that eq.\ (\ref{db})
predicts that the radion gets destabilized at temperature
\beq
\label{Tc}
	T \sim\sqrt{ m_b \Lambda} \sim \epsilon \times {\rm\ TeV}
\eeq
where $\epsilon \sim m_b/\Lambda$ is a small parameter which we shall
specify below.  To understand the dynamics of this phase transition
between compact and noncompact extra dimensions, we must discuss
the mechanism of stabilization.  The most popular such mechanism
was proposed by Goldberger and Wise (GW) \cite{GW}.

\subsection{Goldberger-Wise mechanism}

\begin{figure}[htbp]
\begin{center}
\includegraphics[width=0.45\textwidth,angle=0]{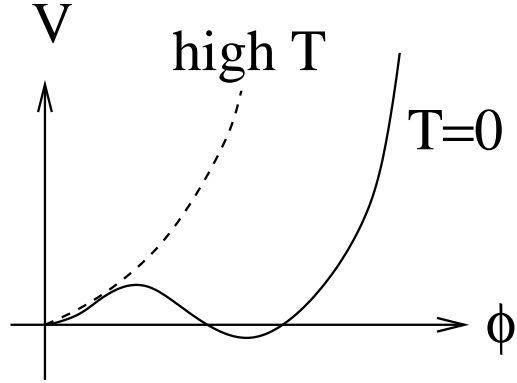}
\end{center}
\caption{The Goldberger-Wise radion potential at zero temperature
(solid curve) and at high temperature (dashed curve).\label{potfig}}
\end{figure}

The GW mechanism uses a bulk scalar field $\Phi$ with Lagrangian
\beq
	{\cal L} = (\partial\Phi)^2 - m^2\Phi^2 - V_0\delta(y)
	- V_1\delta(y-1)
\eeq
where the brane potentials $V_{0,1}$ are minimized at $\Phi = v_0$
and $\Phi= v_1$, respectively.  There are two competing effects in
this model.  The mass term 
\beq
	b\int dy\, e^{-4kby} m^2 \Phi^2
\eeq
is minimized when $b$ is small, while the gradient energy
\beq
	{1\over b}\int dy\, e^{-4kby} \Phi'^2
\eeq
is minimized at large $b$.  The brane potentials insure that the
trivial solution $\Phi=0$ is energetically disfavored; instead 
(assuming the coefficients of $V_{0,1}$ are sufficiently large)
$\Phi$ must vary between the boundary values $v_0$ and $v_1$.
The result of this competition is that $b$ gets stabilized at some
intermediate value,
\beq
\label{bvev}
	b = {1\over k\epsilon} \ln\left({v_0\over v_1}\right) 
\eeq
where $\epsilon \cong m^2/4k^2 \ll 1$.  The canonically normalized
radion field can be written approximately as
\beq
	\phi = f e^{-k b(x)},\quad f = \sqrt{6} M_p
\eeq
and its VEV using (\ref{bvev}) is related to the warp factor by
\beq
	{1\over f}\langle \phi\rangle = e^{-kb} = \eta^{1/\epsilon}	
	\equiv \left({v_1\over v_0}
	\right)^{1/\epsilon}
\eeq
Notice that the large hierarchy is naturally 
generated by raising a moderately
small number $\eta$ to a moderately large power $1/\epsilon$.  
The radion potential has the approximate form
\beq
	V(\phi) = {k v_0^4\over f^4}\phi^4\left[
	\left(\left(\phi\over f\right)^\epsilon-\eta\right)^2
	-\frac{\epsilon}{4} \eta^2\right]
\eeq
This looks like a double-well potential, fig.\ \ref{potfig},
 with a slightly deeper
minimum at $\phi = \langle\phi\rangle$ than at $\phi=0$. (This depth
can be changed by adjusting the TeV brane tension.)  The
local minimum at $\phi=0$ corresponds to the decompactified RS II
theory, where $b\to\infty$. This effective potential predicts 
that the radion mass scales like $\epsilon^{3/4}\Lambda$
\cite{GW2, CF}, but a more
careful analysis shows that actually $m_b \sim \epsilon\Lambda$
\cite{CGK}. 	

\subsection{Decompactification phase transition}
At high temperatures, the GW potential naturally gets a thermal
correction which lifts the nontrivial minimum and leads to a
transition between the compactified and decompactified phases of 
the theory (fig.\ \ref{potfig}).  This was first studied in
\cite{CF}, and has been more recently considered in 
\cite{CNR}-\cite{HR}.
A simple way of understanding the destabilization at high temperature
is through the coupling of the radion to matter.  It couples
conformally, {\it i.e.} to the trace of the stress-energy tensor,
\beq
	{\cal L}_{\rm int} \sim{\phi\over \Lambda} T^\mu_\mu
	= {\phi\over {\rm TeV}} (\rho - 3p)
\eeq
This predicts no shift in the radion during a radiation-dominated era;
however at high $T$ there is also a correction to the radion mass,
$T^2\phi^2$, and its effect is parametrically the same as that of
the linear coupling if we simply estimate $\rho-3p\sim T^4$.  If $V_0$
is the zero-temperature potential, we can estimate the shift in the
radion by taking
\beq
	{dV\over d\phi} = {dV_0\over d\phi} - {T^\mu_\mu\over
\Lambda}=0
\eeq
Letting $\phi = \phi_0 + \delta\phi$, this implies
\beq
	{d^2 V_0\over d\phi^2} \delta\phi  = 
	m^2_b \delta\phi \sim {T^4\over\Lambda}
\eeq
Using the fact that $\phi_0 = f e^{-kb_0} \sim \Lambda$, we see that
\beq
	{\delta\phi\over\phi_0} \sim {T^4\over\Lambda^2 m^2_b}
\eeq
which becomes of order unity at the critical temperature given by
eq.\ (\ref{Tc}). 

More insight into the high-temperature limit of the theory can be
gained by considering the 5D solution to the Einstein
equations at finite temperature: this is the AdS-Schwarzschild
solution where there is a black hole in the bulk, located at
the AdS horizon $r=0$ \cite{Kraus} (which corresponds to $y=\infty$
in the RS coordinate system),
\beq
\label{adsbh}
	ds^2 = -\left({r^2\over \ell^2}-{\mu\over r^2}\right)dt^2
	+r^2 d\vec x^{\,2} + \left({r^2\over \ell^2}-{\mu\over
	r^2}\right)^{-1}dr^2
\eeq  
The temperature of the system is identified with the Hawking
temperature of the black hole, whose mass is $\mu$, and 
$\ell= 1/k$ is the AdS curvature length scale.  
In this coordinate system the branes are moving through the bulk,
toward larger values of $r$, with the Planck brane at $r = R(t)$
say, and the TeV brane at some smaller value of $r$.  There is a
horizon at
\beq
	r = \mu^{1/4}\ell
\eeq
behind which the TeV brane is hidden before the compactification phase
transition. 

The brane motion can be understood by considering the induced metric
on the brane at $r=R(t)$:
\beq
	ds^2 = -\left({R^2\over \ell^2}-{\mu\over R^2}
	- \dot R^2 \left({R^2\over \ell^2}-{\mu\over
	R^2}\right)^{-1}\right)dt^2 - R^2 d\vec x^{\,2}
	\equiv -d\tau^2 + R^2(\tau) d\vec x^{\,2}
\eeq
where $\tau$ is the proper time for an observer on the brane.
It is clear from the latter form that $R(\tau)$ plays the role of the
scale factor for the cosmological expansion measured by the brane
observer.  The dynamics of $R(\tau)$ are determined by the junction
condition at the brane, leading to the Friedmann equation 
\beq
	H^2 = {\mu\over R^4}
\eeq
which has the interpretation of $\mu/R^4$ being the energy
density of ``dark radiation,'' the Hawking radiation emitted
by the black hole, consisting of thermal KK gravitons in the bulk.
Notice that this Friedmann equation has the normal form, $H^2\sim
\rho$, rather than $\rho^2$, since radiation redshifts like $1/R^4$.
In the AdS/CFT correspondence, these are the thermalized CFT degrees of
freedom.  This picture in which the geometry is static but the branes are
moving has been called ``mirage cosmology,'' \cite{mirage} 
since the expansion of
the universe appears to be an illusion on the part of the brane
observers, due to their motion through the bulk.

Of course, the same physics can also be seen in the RS coordinates,
where the geometry is explicitly time dependent and the branes are
stationary \cite{BDEL}.  The metric functions are
\beq
	a^2(t,y) = a_0^2(t) e^{-2kby} + {c\over a_0^2(t)}
	\sinh^2 kby; \qquad n(t,y) = {a_0\over a}
	\left( e^{-2kby} - {c\over a_0^4(t)} \sinh^2 kby\right)
\eeq
and the horizon is at
\beqa
	{e^{-2kby}\over \sinh^2 kby} &\cong& 4e^{-4kby} = 
	{c\over a_0^4}
	\nonumber\\ &\Longrightarrow& y = {1\over 4kb}\ln {4c\over a_0^4}
\eeqa
which recedes toward the AdS horizon as the universe expands.  
In these coordinates $c/a_0^4$ is the dark radiation density, so
$c\sim\mu$ in terms of the coordinates in (\ref{adsbh}).  At some
critical temperature the TeV brane is uncloaked by the horizon,
signaling the compactification phase transition \cite{AHPR}.  In the
prior epoch when the TeV brane is hidden below the horizon, it is as
good as nonexistent, since it is inaccessible to observers outside the
horizon.  For them, the TeV brane might as well be at $r=0$, which is
the decompactified phase of the theory. 

The purely geometrical interpretation given above does not include
the effects of radion stabilization.  From the shape of the radion
potential shown in fig.\ \ref{potfig}, we see that there is a
first-order phase transition of a peculiar kind: within the high-$T$
5D single-brane universe, bubbles nucleate, inside of which the
universe is 4D and the TeV brane has appeared.  References
\cite{CF}-\cite{RS} show that the phase transition does not
necessarily complete (as in old inflation) for some ranges of
parameters of the GW stabilization mechanism.  Ref.\ \cite{CF}
found that the bubble nucleation rate $\Gamma \sim e^{-S}$ is
suppressed by the tunneling action
\beq
	S\sim {\Lambda^2 M_p^2 e^{-2/\sqrt{\epsilon}}\over
	\epsilon v_1 \sqrt{k} T^2} 
\sim {e^{-2/\sqrt{\epsilon}}\Lambda^3\over T^2 m_b} 
\eeq
at temperature $T$, so the transition is slower if the radion mass
$m_b$ is small or if $\epsilon$ is large.  Ref.\ \cite{CNR} obtains
quantitatively different but qualitatively similar results.

An interesting recent development is the prediction of gravity waves
produced during the phase transition, which can be large enough to be
observable at LISA if the phase transition is strongly enough first
order \cite{RS}.  The gravity waves are produced by bubble collisions
and by turbulence.  However they are only observably large for model
parameters where perturbation theory is starting to break down, so it
would be interesting to undertake a more careful calculation of the
effect.  

\section{The DGP model}
Dvali, Gabadadze and Porrati (DGP) found a remarkably simple way in
which branes can modify gravity at large distances \cite{DGP},
using the action
\beq
	S = 2M_p^2\int d^{\,4}x\,\sqrt{-g}\, R^{(4)} 
	+ 2M_5^3\int d^{\,4}x\,dy\, \sqrt{-G}\, R^{(5)} 
\eeq
where $g$ and $G$ is are the 4D and 5D metric determinants,
respectively.  By including the 4D Einstein term on a 3-brane, 
gravity is quasilocalized there, up to the distance scale
\beq
	r_c \sim {M_p^2\over M_5^3}
\eeq
At distances $r < r_c$, gravity looks 4D, while for $r> r_c$ it looks
5D.  The cosmology of this model was first studied in \cite{Def,DDG},
which found the modified Friedmann equation
\beq
	{\rho\over 3 M_p^2} = H^2 \pm {H\over r_c}
\eeq
having two branches of solutions,
\beq
	H = \sqrt{{\rho\over 3 M_p^2}+{1\over 4 r_c^2}}\mp
	{1\over 2 r_c}
\eeq
The upper sign leads to conventional cosmology as $\rho\to 0$,
while the lower one has the interesting property that $H\to 1/r_c$
even as the matter density on the brane vanishes.  This is called
the {\it self-accelerating} solution, and gives rise to the possibility
of explaining the accelerated expansion of the universe without an 
explicit cosmological constant term.  To match the current Hubble
rate, a small 5D gravity scale is required:
\beq
	M_5 = (M_p^2 H_0)^{1/3} \sim 10 {\rm \ MeV}
\eeq

In principle this kind of acceleration is experimentally
distinguishable from that due to dark energy, even if it has
an arbitrary (but constant) equation of state $w = p/\rho$.
The luminosity distance $d_L(z)$ as a function of redshift $z$,
\beq
	d_L = (1+z)\int_0^z dx {H_0\over H(x)}
\eeq
has a different shape than in conventional dark energy models,
so if the expansion history can be probed accurately as a function of
$z$, as with type I supernovae, one can test the DGP model relative
to fluid models.  

\begin{figure}[htbp]
\begin{center}
\includegraphics[width=0.45\textwidth,angle=0]{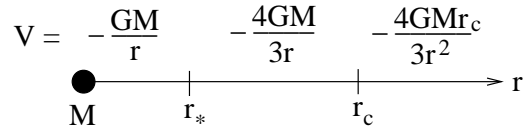}
\end{center}
\caption{Distance scales and corresponding behavior of gravitational
potential in the DGP model.\label{dgpfig}}
\end{figure}

Interestingly, the DGP model also predicts nonstandard gravity at
much shorter scales, potentially including the solar system.  This
comes about because of an extra scalar graviton polarization $\pi$
which changes the strength of the Newtonian gravitational potential
in a distance-dependent way.  One can think of $\pi$ as a
``brane-bending'' mode, associated with transverse fluctuations of
the brane.  It modifies the strength of gravity at distances $r >
r_*= (r_g r_c^2)^{1/3}$, where $r_g=2M/M_p^2$ is the gravitational
radius  of a source of mass $M$, for example the sun if we are the
testing solar system.  Thus
\beq
	r_* = \left({2M M_p^2\over M_5^3}\right)^{1/3}
\eeq
Naively, the extra polarization $\pi$ would spoil GR down to zero
distance, but nonlinear terms in the Einstein equations become
important when $r< r_*$ and these suppress the scalar contribution to
the gravitational potential \cite{DDGV}.  The different distance
regimes and corresponding gravitational potentials are illustrated 
in fig.\ \ref{dgpfig}.

Of course even in the 4D regime, there are small corrections to the
Newtonian potential, and these lead to anomalous precession of
planetary orbits, and that of the moon.  The precession rate turns out
to be independent of the masses of the source or the orbiting body
(see \cite{Lue}),
\beq
	{d\Delta\phi\over dt} = \mp{3\over 8 r_0} = \mp\,
	5\mu{\rm as/year}
\eeq
which might be probed by future lunar laser ranging experiments.

There is yet another mass/distance scale associated with the $\pi$ 
mode,
\beq
	\Lambda = {M_5^2\over M_p}\equiv r_\Lambda^{-1}\sim
	(10^3{\rm\ km})^{-1}
\eeq
For $r< r_\Lambda$, the self coupling
$(\partial\pi)^2\square\pi/\Lambda^3$ of the effective Lagrangian
for  the $\pi$ becomes strong.  If generic operators of the form
$(\partial\pi)^{2N}/\Lambda^{4N-4}$ were also present, as would be
true for an ordinary Goldstone boson, this strong coupling would
cause the theory to lose predictivity at any $r< r_\Lambda$. 
Remarkably, such operators are not present in the DGP model, due to
the unconventional shift symmetry $\partial_\mu\pi \to
\partial_\mu\pi + c_\mu$, which is a remnant of the 5D Lorentz
symmetry.  Hence predictivity is not spoiled in the DGP model. 
However, precisely the absence of the $(\partial\pi)^{4}/\Lambda^{4}$
term leads to a violation of causality and locality, as shown by
ref.\ \cite{AADNR}.  This indicates that the DGP model has no
ultraviolet completion which respects analyticity of the S-matrix,
hence making it appear quite unlikely that it can be embedded within
string theory.


\begin{thebibliography}{99}
\bibitem{ADD}
  N.~Arkani-Hamed, S.~Dimopoulos and G.~R.~Dvali,
  ``Phenomenology, astrophysics and cosmology of theories with  sub-millimeter
  dimensions and TeV scale quantum gravity,''
  Phys.\ Rev.\  D {\bf 59}, 086004 (1999)
  [arXiv:hep-ph/9807344].

\bibitem{BDL}
  P.~Binetruy, C.~Deffayet and D.~Langlois,
  ``Non-conventional cosmology from a brane-universe,''
  Nucl.\ Phys.\  B {\bf 565}, 269 (2000)
  [arXiv:hep-th/9905012].

\bibitem{HW}
  P.~Ho\v rava and E.~Witten,
  ``Eleven-Dimensional Supergravity on a Manifold with Boundary,''
  Nucl.\ Phys.\  B {\bf 475}, 94 (1996)
  [arXiv:hep-th/9603142].

\bibitem{CGKT}
  C.~Csaki, M.~Graesser, C.~F.~Kolda and J.~Terning,
  ``Cosmology of one extra dimension with localized gravity,''
  Phys.\ Lett.\  B {\bf 462}, 34 (1999)
  [arXiv:hep-ph/9906513].


\bibitem{CGS}
  J.~M.~Cline, C.~Grojean and G.~Servant,
  ``Cosmological expansion in the presence of an extra dimension,''
  Phys.\ Rev.\ Lett.\  {\bf 83}, 4245 (1999)
  [arXiv:hep-ph/9906523].

\bibitem{RSI}
  L.~Randall and R.~Sundrum,
  ``A large mass hierarchy from a small extra dimension,''
  Phys.\ Rev.\ Lett.\  {\bf 83}, 3370 (1999)
  [arXiv:hep-ph/9905221].

\bibitem{Kaloper}
  N.~Kaloper,
  ``Bent domain walls as braneworlds,''
  Phys.\ Rev.\  D {\bf 60}, 123506 (1999)
  [arXiv:hep-th/9905210].

\bibitem{CGRT}
  C.~Csaki, M.~Graesser, L.~Randall and J.~Terning,
  ``Cosmology of brane models with radion stabilization,''
  Phys.\ Rev.\  D {\bf 62}, 045015 (2000)
  [arXiv:hep-ph/9911406].

\bibitem{CV}
  J.~M.~Cline and J.~Vinet,
  ``Order $\rho^2$ corrections to Randall-Sundrum I cosmology,''
  JHEP {\bf 0202}, 042 (2002)
  [arXiv:hep-th/0201041].

\bibitem{RSII}
  L.~Randall and R.~Sundrum,
  ``An alternative to compactification,''
  Phys.\ Rev.\ Lett.\  {\bf 83}, 4690 (1999)
  [arXiv:hep-th/9906064].

\bibitem{eotwash}
  D.~J.~Kapner, T.~S.~Cook, E.~G.~Adelberger, J.~H.~Gundlach, B.~R.~Heckel, C.~D.~Hoyle and H.~E.~Swanson,
  ``Tests of the gravitational inverse-square law below the dark-energy length
  scale,''
  Phys.\ Rev.\ Lett.\  {\bf 98}, 021101 (2007)
  [arXiv:hep-ph/0611184].

\bibitem{MWBH}
  R.~Maartens, D.~Wands, B.~A.~Bassett and I.~Heard,
  ``Chaotic inflation on the brane,''
  Phys.\ Rev.\  D {\bf 62}, 041301 (2000)
  [arXiv:hep-ph/9912464].

\bibitem{LS}
  A.~R.~Liddle and A.~J.~Smith,
  ``Observational constraints on braneworld chaotic inflation,''
  Phys.\ Rev.\  D {\bf 68}, 061301 (2003)
  [arXiv:astro-ph/0307017].

\bibitem{CLL}
  E.~J.~Copeland, A.~R.~Liddle and J.~E.~Lidsey,
  ``Steep inflation: Ending braneworld inflation by gravitational particle
  production,''
  Phys.\ Rev.\  D {\bf 64}, 023509 (2001)
  [arXiv:astro-ph/0006421].

\bibitem{GW}
  W.~D.~Goldberger and M.~B.~Wise,
  ``Modulus stabilization with bulk fields,''
  Phys.\ Rev.\ Lett.\  {\bf 83}, 4922 (1999)
  [arXiv:hep-ph/9907447].

\bibitem{GW2}
  W.~D.~Goldberger and M.~B.~Wise,
  ``Phenomenology of a stabilized modulus,''
  Phys.\ Lett.\  B {\bf 475}, 275 (2000)
  [arXiv:hep-ph/9911457].


\bibitem{CF}
  J.~M.~Cline and H.~Firouzjahi,
  ``Brane-world cosmology of modulus stabilization with a bulk scalar  field,''
  Phys.\ Rev.\  D {\bf 64}, 023505 (2001)
  [arXiv:hep-ph/0005235].


\bibitem{CGK}
  C.~Csaki, M.~L.~Graesser and G.~D.~Kribs,
  ``Radion dynamics and electroweak physics,''
  Phys.\ Rev.\  D {\bf 63}, 065002 (2001)
  [arXiv:hep-th/0008151].

\bibitem{CNR}
  P.~Creminelli, A.~Nicolis and R.~Rattazzi,
  ``Holography and the electroweak phase transition,''
  JHEP {\bf 0203}, 051 (2002)
  [arXiv:hep-th/0107141].

\bibitem{RS}
  L.~Randall and G.~Servant,
  ``Gravitational waves from warped spacetime,''
  arXiv:hep-ph/0607158.

\bibitem{HR}
  G.~T.~Horowitz and M.~M.~Roberts,
  ``Dynamics of first order transitions with gravity duals,''
  JHEP {\bf 0702}, 076 (2007)
  [arXiv:hep-th/0701099].


\bibitem{Kraus}
  P.~Kraus,
  ``Dynamics of anti-de Sitter domain walls,''
  JHEP {\bf 9912}, 011 (1999)
  [arXiv:hep-th/9910149].

\bibitem{mirage}
  A.~Kehagias and E.~Kiritsis,
  ``Mirage cosmology,''
  JHEP {\bf 9911}, 022 (1999)
  [arXiv:hep-th/9910174].

\bibitem{BDEL}
  P.~Binetruy, C.~Deffayet, U.~Ellwanger and D.~Langlois,
  ``Brane cosmological evolution in a bulk with cosmological constant,''
  Phys.\ Lett.\  B {\bf 477}, 285 (2000)
  [arXiv:hep-th/9910219].

\bibitem{AHPR}
  N.~Arkani-Hamed, M.~Porrati and L.~Randall,
  ``Holography and phenomenology,''
  JHEP {\bf 0108}, 017 (2001)
  [arXiv:hep-th/0012148].

\bibitem{DGP}
  G.~R.~Dvali, G.~Gabadadze and M.~Porrati,
  ``4D gravity on a brane in 5D Minkowski space,''
  Phys.\ Lett.\  B {\bf 485}, 208 (2000)
  [arXiv:hep-th/0005016].

\bibitem{Def}
  C.~Deffayet,
  ``Cosmology on a brane in Minkowski bulk,''
  Phys.\ Lett.\  B {\bf 502}, 199 (2001)
  [arXiv:hep-th/0010186].

\bibitem{DDG}
  C.~Deffayet, G.~R.~Dvali and G.~Gabadadze,
  ``Accelerated universe from gravity leaking to extra dimensions,''
  Phys.\ Rev.\  D {\bf 65}, 044023 (2002)
  [arXiv:astro-ph/0105068].

\bibitem{DDGV}
  C.~Deffayet, G.~R.~Dvali, G.~Gabadadze and A.~I.~Vainshtein,
  ``Nonperturbative continuity in graviton mass versus perturbative
  discontinuity,''
  Phys.\ Rev.\  D {\bf 65}, 044026 (2002)
  [arXiv:hep-th/0106001].

\bibitem{Lue}
  A.~Lue,
  ``The phenomenology of Dvali-Gabadadze-Porrati cosmologies,''
  Phys.\ Rept.\  {\bf 423}, 1 (2006)
  [arXiv:astro-ph/0510068].

\bibitem{AADNR}
  A.~Adams, N.~Arkani-Hamed, S.~Dubovsky, A.~Nicolis and R.~Rattazzi,
  ``Causality, analyticity and an IR obstruction to UV completion,''
  JHEP {\bf 0610}, 014 (2006)
  [arXiv:hep-th/0602178].

\end{thebibliography}
\end{document}